\documentclass[aps,pra,eqsecnum,twocolumn,superscriptaddress]{revtex4}
\usepackage{graphicx}
\usepackage{bm}
\usepackage{amsmath,amssymb,amsthm,dsfont,bm}
\usepackage{color}
\usepackage{wasysym}
\usepackage{ulem}
\usepackage{appendix}
\usepackage[dvipsnames]{xcolor}

\usepackage[english]{babel} 
\usepackage[utf8]{inputenc} 
\usepackage{hyperref}
\usepackage{cancel}

\begin{document}

\title{Coherence via reiterated beam splitting}
\author{Guillermo D\'iez}
\affiliation{Departamento de \'Optica, Facultad de Ciencias
F\'{\i}sicas, Universidad Complutense, 28040 Madrid, Spain}
\author{Laura Ares}
\affiliation{Departamento de \'Optica, Facultad de Ciencias
F\'{\i}sicas, Universidad Complutense, 28040 Madrid, Spain}
\affiliation{Theoretical Quantum Science, Institute for Photonic Quantum Systems (PhoQS), Paderborn University, Warburger Stra\ss{}e 100, 33098 Paderborn, Germany}
\author{Alfredo Luis}
\affiliation{Departamento de \'Optica, Facultad de Ciencias
F\'{\i}sicas, Universidad Complutense, 28040 Madrid, Spain}
\date{\today}

\begin{abstract}
Beam splitters are not-free operations with regard to quantum coherence. As a consequence, they can create coherence from both coherent and incoherent states. We investigate the increase in coherence produced by cascades of beam splitters. To this end, we construct two different configurations and analyze different sequences of input states. 
\end{abstract}
\maketitle

\section{Introduction} 

The beam splitter is a basic device in optics conceived to create coherence in the classical realm and is therefore a fundamental element in interferometry. Its role is no less relevant in quantum optics \cite{MW95,LS95}. In this realm, it allows one to easily perform fundamental quantum state transformations, such as entanglement generation, state measurements like tomography, and certification of nonclassicality. For this reason, it is present in most photonic implementations of quantum applications \cite{logic,BSampling,DNM21}. Thus, we investigate beam splitting as a coherence-making process in quantum optics.

In recent years, we have witnessed a rapid growth in interest in quantum coherence as the cornerstone of quantum theory \cite{JA06,BCP14,SP17}. Far from the more standard theory of coherence in quantum optics, based on the works of Glauber and Sudarshan \cite{RG63a,RG63b,ECGS63}, current formulations of quantum coherence mainly focus on its resourcefulness as a quantifier of quantum superpositions \cite{CG19, SP17, AW16}. This novel approach is definitely more versatile within the ongoing technological perspective, which motivates the effort to translate the theoretical framework to the laboratory \cite{WSRXCG21,SY22}. It has been shown that quantum coherence improves the efficiency of Grover's, Deutsch-Jozsa's and Shor's algorithms \cite{SLWYYF17,H16,ATEMP22}; it is directly related to path-information uncertainty \cite{BBCH16, BQSP15} and purity \cite{CH15}, and it plays an important role in quantum thermodynamics \cite{SCLP19}, and condensed-matter scenarios \cite{JS23}. However, behind the operational perspective, a lot of theoretical research is still expanding the resource theory of quantum coherence \cite{BSFPW17,RFWA18}. For example, the concepts of distillation, dilution, and catalysis relate to the fundamental question of defining the set of free operations \cite{ZLYCW19,DKMS23}.

In this regard, beam splitters are not-free operations even in the more restrictive contexts in which the inability to use coherence is taken into account \cite{YGGV16}. Thus, we face the question of how much coherence can be introduced by beam splitting. In the context of quantum correlations, this question has usually focused on the generation of entanglement, which excludes the option of Glauber coherent input states \cite{MSK02}. However, Glauber coherent states have proven to be a useful tool for quantum tasks \cite{TF16}, and quantum coherence goes beyond entanglement, so we broaden the point of view for the analysis.

In this work we look for the optimal configuration of a set of beam splitters in order to maximize the quantum coherence of the output state. The analysis is grounded on the study of an individual beam splitter as a coherence maker introduced in Ref. \cite{AL23}. This is especially motivated by the rather unsettling possibility of unlimited coherence growth with an unlimited increase in the number of beam splitters. We especially focus on the mode decomposition of the total coherence, trying to relate the overall coherence to the coherence of each output mode.

\bigskip

Throughout, we consider lossless beam splitters and compute coherence in the photon-number basis $|n \rangle$. As a suitable coherence measure, we utilize the $l_1$-norm of coherence \cite{BCP14,SP17},
\begin{equation}
\label{Cl1}
 \mathcal{C} = \sum_{n,n^\prime}| \langle n | \rho |n^\prime \rangle | -1,
\end{equation}
where $\rho$ is the density matrix. It simplifies for pure states, $ |\psi \rangle = \sum_{n} c_n | n \rangle $, into the form 
\begin{equation}
\label{fps}
   \quad \mathcal{C} = \left ( \sum_{n} |c_n |\right )^2 -1,
\end{equation}
where $n$ represents any collection of natural numbers needed to label the photon-number basis in a multi-mode scenario. 

\bigskip

The state acting as the input of the series of beam splitters is designed such that only one input mode is populated, while all the other input modes are in vacuum states. This is the traditional way in which beam splitting is used to manage coherence. For the field state in the populated mode, we examine pure and mixed states, both coherent and incoherent. These are Glauber coherent states $|\alpha \rangle$ as pure and partially coherent inputs, number states $| n\rangle$ as pure and incoherent inputs, and, finally, phase-averaged and thermal states as both mixed and incoherent inputs. With these choices, we examine whether the increase in coherence depends on the initial coherence conveyed by the input field state, its quantumness, or purity. 

\section{Coherent-state input}

\begin{figure}[h]
    \includegraphics[width=8cm]{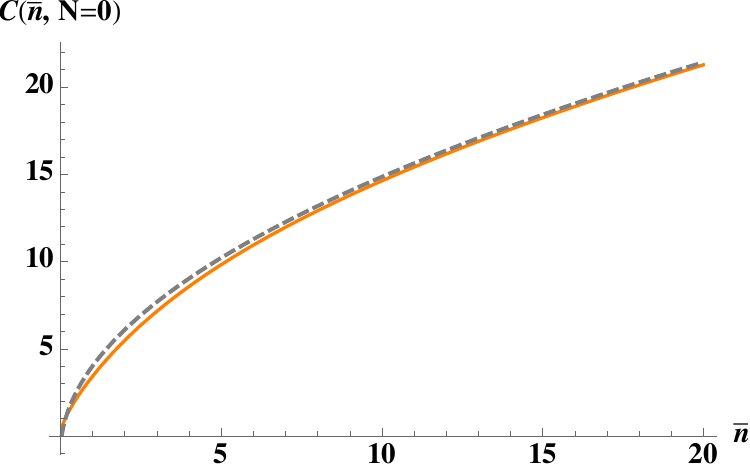}
    \caption{Plot of $\mathcal{C} (\bar{n}, N=0 )$ for a single coherent state as a function of the total mean number of photons $\bar{n}$ (orange solid line) and its approximation in Eq. \ref{Ccoheapp} (gray dashed line).}
    \label{cohe}
\end{figure}

First, we consider a Glauber coherent state $|\alpha \rangle$ in the only populated mode, so that the input state becomes 
\begin{equation}
  |\alpha ,0,\ldots,0 \rangle.  
\end{equation}
Let us compute the coherence for the input state. The coefficients in the single-mode photon-number basis have modulus
\begin{equation}
\label{cn}
| c_n | = \sqrt{\frac{\bar{n}^n}{n!}} e^{-\bar{n}/2},
\end{equation}
where $\bar{n}$ is the mean number of photons $\bar{n} = |\alpha |^2$. The coherence of the input state, which we shall denote $\mathcal{C} (\bar{n}, N=0 )$, where $N$ will later represent the number of beam splitters, is 
\begin{equation}
\label{Cc}
    \mathcal{C} (\bar{n}, N=0 ) = e^{-\bar{n}}\left ( \sum_{n=0}^\infty \sqrt{\frac{\bar{n}^n}{n!}}  \right )^2-1 .
\end{equation}
In order to obtain the intuition of analytical solutions we utilize the following approximation: For large enough $\bar{n}$, the sum in Eq. \ref{Cc} can be approximated by an integral, and the Poissonian statistics in Eq. \ref{cn} can be approximated by a Gaussian, 
\begin{equation}
\label{Posi}
| c_n | \simeq \frac{1}{(2 \pi \bar{n})^{1/4}} e^{-(n-\bar{n})^2/(4 \bar{n})}.
\end{equation}
Under this assumption, the coherence results in
\begin{equation}
\label{Ccoheapp}
    \mathcal{C} (\bar{n}, N=0 ) \simeq 2\sqrt{2 \pi \bar{n}}-1 .
\end{equation}
In Fig. \ref{cohe}, we plot $\mathcal{C} (\bar{n}, N=0 )$ for a single coherent state as a function of the total mean number of photons $\bar{n}$ (orange solid line) and its approximation in Eq. (\ref{Ccoheapp}) (gray dashed line). 

{}


It is worth noting the increase of coherence with the mean number of photons, without an upper limit \cite{ZSLF16}. This behavior is not contemplated in the Glauber-Sudarshan quantum-optical theory of coherence or in the classical theory of coherence, where coherence is independent of the field intensity. This independence, by construction, reflects the idea that coherence is a measure of the wavelike quality independent of the amount of light. 

\bigskip

\begin{figure}[h]
    \includegraphics[width=6.5cm]{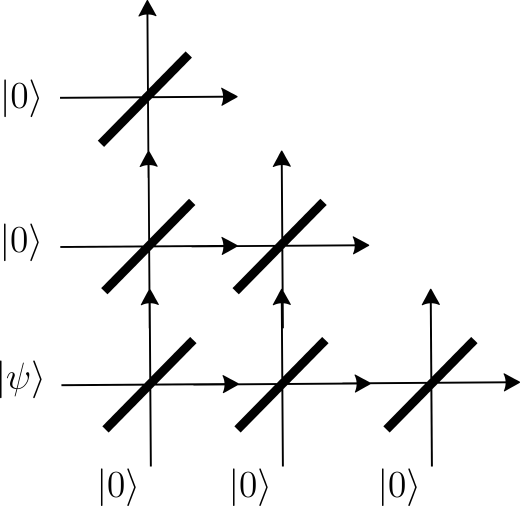}
    \caption{Configuration 1.  }
    \label{cascada1}
\end{figure}

\begin{figure}[h]
    \includegraphics[width=6cm]{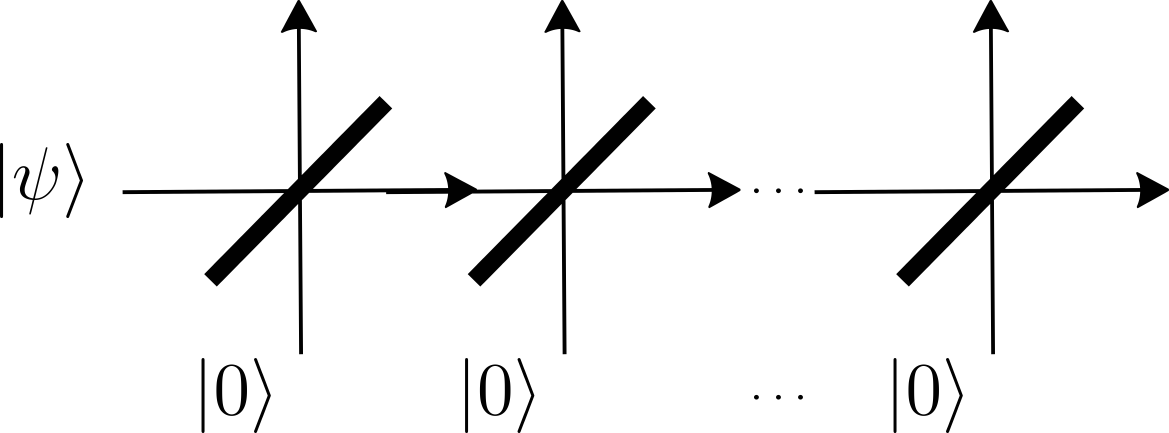}
    \caption{Configuration 2.  }
    \label{cascada2}
\end{figure}

Now, we can address the coherence after a cascade of $N$ arbitrary lossless beam splitters, such as the ones illustrated in Figs. \ref{cascada1} and \ref{cascada2}. We take advantage of the known properties of coherent states and beam splitters to note that the output state will be a product of coherent states in all the $N+1$ output modes, that is 
\begin{equation}
     |\beta_0 \rangle | \beta_1 \rangle \ldots |\beta_N \rangle ,
\end{equation}
where $\beta_j = \tau_j \alpha$, with $\tau_j$ being the transmission coefficients linking the complex amplitude of the corresponding output mode with the complex amplitude of the coherent state in the input mode. For example, for configuration 2 in Fig. \ref{cascada2} we have 
\begin{equation}
    \tau_0 = t_1\cdots t_N, \quad \tau_1 = r_1, \quad \tau_{j>1} = t_1 \cdots t_{j-1} r_j ,
\end{equation}
for $j=1, \ldots, N$, where $t_j$ and $r_j$ are the corresponding  transmission and reflection coefficients of each beam splitter numbered from left to right in Fig. \ref{cascada2}. 

Thanks to this factorization, the total coherence at the output can be expressed in the form
\begin{equation}
\label{CT}
 \mathcal{C} (\bar{n}, N) =  \Pi_{j=0}^N\left [ \mathcal{C} (\bar{n}_j, 0)+1 \right ]-1,
 \end{equation}
 where $\mathcal{C} (\bar{n}_j, 0)$ is the coherence of a single-mode coherent state in Eq. (\ref{Cc}), and $\bar{n}_j = |\beta_j|^2$ is the mean number of photons of the coherent state in the corresponding output mode, with 
 \begin{equation}
 \label{ec}
     \bar{n} = \sum_{j=0}^N \bar{n}_j ,
 \end{equation}
 by energy conservation. 

{}
{}
\bigskip

In order to characterize the set of beam splitters as a coherence maker entity, we look for the optimum choice of parameters for each individual beam splitter, so that $\mathcal{C} (\bar{n}, N)$ in Eq. (\ref{CT}) becomes the maximum. Thanks to the factorized form in Eq. (\ref{CT}), we can easily demonstrate the following theorem: The maximum of $\mathcal{C} (\bar{n}, N)$ in Eq. (\ref{CT}) occurs when the mean photon numbers of all output modes are equal, 
 \begin{equation}
 \label{mpne}
     \bar{n}_j = \frac{\bar{n}}{N+1} ,
 \end{equation}
and then, 
\begin{equation}
\label{TC}
    \mathcal{C}_{\rm max} (\bar{n}, N) = \left [   \mathcal{C} \left ( \frac{\bar{n}}{N+1} , 0 \right ) +1 \right ]^{N+1} - 1 .
\end{equation}
After the Gaussian approximation in Eq. (\ref{Ccoheapp}), which we have shown works even for a small mean number of photons, the coherence becomes
\begin{equation}
\label{CTapp}
\mathcal{C}_{\rm max} (\bar{n}, N ) \simeq \left ( 8 \pi \frac{\bar{n}}{N+1} \right )^{\frac{N+1}{2}}-1 .
\end{equation}

\bigskip

Let us address the demonstration of the theorem. For large mean photon numbers, where Eq. (\ref{Ccoheapp}) holds, a simple proof is available since in such a case the state-dependent term in Eq. (\ref{CT}) is proportional to 
\begin{equation}
 \Pi_{j=0}^N \sqrt{\bar{n}_j} .
\end{equation}
Taking into account energy conservation in Eq. (\ref{ec}), the maximum of this factor via Lagrange multipliers leads to a unique extreme point reached when the equipartition of energy [Eq. (\ref{mpne})] holds, which is clearly the maximum.

\bigskip

In the general case, the theorem can be demonstrated without approximations by {\it reductio ad absurdum}. Let us assume that the maximum coherence $\mathcal{C}_{\rm max}$ holds when at least two of the mean photon numbers are not equal, say $\bar{n}_0 \neq \bar{n}_1$ without loss of generality. We can focus on the contribution of these modes to the total coherence 
\begin{equation}
\label{fact}
\left [ \mathcal{C} (\bar{n}_0, 0)+1 \right ] \left [\mathcal{C} (\bar{n}_1, 0)+1 \right ] .
\end{equation}
We can show that this contribution can be increased by an equal distribution of the $\bar{n}_0 + \bar{n}_1$ photons between the two output modes without affecting the rest, contradicting the assumed maximum. 

\bigskip

\begin{figure}[h]
    \includegraphics[width=6.5cm]{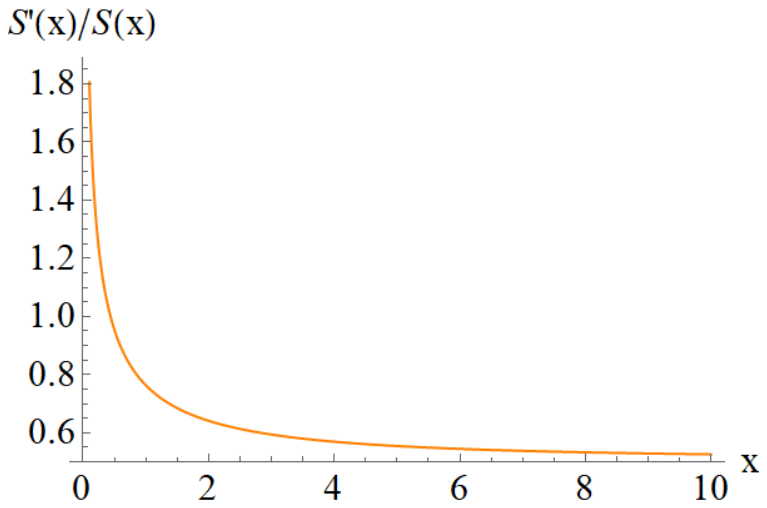}
    \caption{Plot of the function $S^\prime(x)/S(x)$, showing its monotone character.}
    \label{mono}
\end{figure}

To this end the contribution in Eq. (\ref{fact}) depending on the splitting of the photons restricted to a constant total mean number of photons shared by the two modes, say, $\bar{n}_{0,1}=\bar{n}_0 + \bar{n}_1$, can be expressed as
\begin{equation}
    S(x)S(\bar{n}_{0,1}-x) ,
\end{equation}
where
\begin{equation}
\label{dS1}
    S(x) = \sum_{n=0}^\infty \sqrt{\frac{x^n}{n!}} 
\end{equation}
and $x$ represents the number of photons in one of the output modes, say $x= \bar{n}_0$. The standard procedure to look for extremes leads to satisfaction of the equation 
\begin{equation}
\label{SpS}
    \frac{S^\prime(x)}{S(x)}= \frac{S^\prime(\bar{n}_{0,1}-x)}{S(\bar{n}_{0,1}-x)}  ,
\end{equation}
where the prime denotes the derivative with respect to $x$. It is clear that the equal splitting $x=\bar{n}_{0,1}/2$ is a solution. That this is unique can be inferred from the fact that the function 
\begin{equation}
    \frac{S^\prime(x)}{S(x)}= \frac{1}{2}\frac{\sum_{n=0}^\infty n \sqrt{x^{n-2}/n!}}{\sum_{n=0}^\infty \sqrt{x^n/n!}}
\end{equation} 
is monotone, as shown in Fig. \ref{mono}. Then uniqueness follows because the two sides of Eq. (\ref{SpS}) are mutual mirror images with respect their dependence on $x$. That the extreme is the maximum follows from the fact that coherence is non-negative and for both extremes $x=0, \bar{n}_{0,1}$ the coherence vanishes. 

{}

\bigskip

Intuitively, the theorem can be understood from the fact that there is full symmetry under the exchange between transmission and reflection so their equality must be an extreme. Such an extreme must be the maximum since the spreading of the photon-number distribution is clearly largest at 50\% splitting, and the larger the spreading is, the larger the coherence is.

\begin{figure}[h]
    \includegraphics[width=8cm]{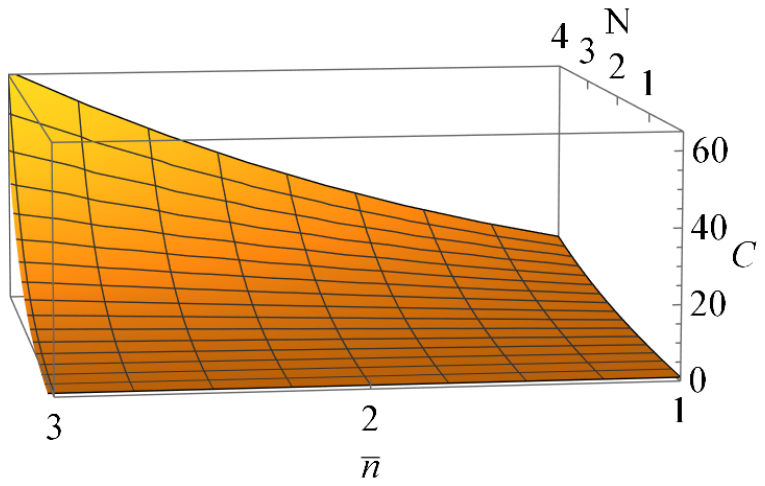}
    \caption{Plot of the quotient $\mathcal{C} = \mathcal{C}_{\rm max} (\bar{n}, N ) /\mathcal{C}_{\rm max} (\bar{n}, 0 )$ as a function of the number of beam splitters $N$ and the mean number of photons, $\bar{n}$, showing a clear increase in coherence with both the number of photons and the number of beam splitters.}
    \label{coheN}
\end{figure}

In Fig. \ref{coheN}, we plot the gain in coherence caused by the reiteration of splitting,  $\mathcal{C} = \mathcal{C}_{\rm max} (\bar{n}, N ) /\mathcal{C}_{\rm max} (\bar{n}, 0 )$ as a function of the number of beam splitters $N$ and the mean number of photons $\bar{n}$. A clear increase in coherence can be appreciated with regard to the increase in both the number of photons and the number of beam splitters.

{}

\bigskip

To go beyond numerical results that may be valid for a small range of physical parameters, let us demonstrate analytically the limitless increase in coherence with an increasing number of beam splitters $N$ for a fixed mean number of photons $\bar{n}$. To this end we note that approximation (\ref{CTapp}) might fail since for large $N$ the mean number of photons in each coherent state tends to be so small that the Gaussian approximation of the photon-number distribution is no longer valid. 

\bigskip

We can prove the increase in coherence with $N$ by deriving a suitable lower bound to $\mathcal{C}_{\rm max} (\bar{n}, N )$. To this end we note that after Eqs. (\ref{Cc}) and (\ref{TC}) 
\begin{equation}
\label{}
    \mathcal{C}_{\rm max} (\bar{n}, N )  = e^{-\bar{n}} \left [S \left (\frac{\bar{n}}{N+1} \right ) \right ]^{2(N+1)}-1
\end{equation}
where $S(x)$ was already defined in Eq. (\ref{dS1}). Since $n! \geq \sqrt{n!}$, we get
\begin{equation}
\label{dS}
    S(x) = \sum_{n=0}^\infty \sqrt{\frac{x^n}{n!}} \geq  \sum_{n=0}^\infty \frac{\sqrt{x^n}}{n!} = e^{\sqrt{x}},
\end{equation}
and then 
\begin{equation}
\label{lwG}
    \mathcal{C}_{\rm max} (\bar{n}, N ) \geq e^{-\bar{n}} e^{2 \sqrt{\bar{n}(N+1)}} -1 ,
\end{equation}
which proves the limitless increase in coherence with an increasing number of beam splitters $N$.

\bigskip

\begin{figure}[h]
    \includegraphics[width=8cm]{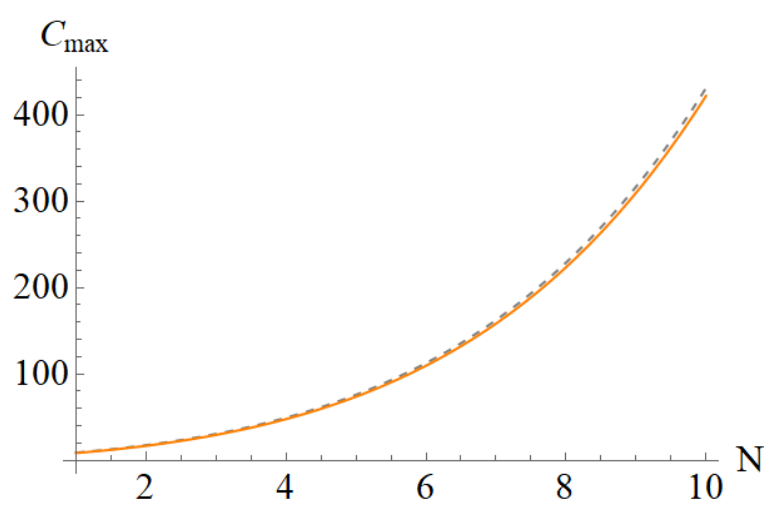}
    \caption{Comparison of the asymptotic behavior in Eq. (\ref{Casymp}) (orange solid line) with a numerical evaluation of Eq. (\ref{TC}) (gray dashed line) for $\bar{n}=1$ and $N$ ranging from 1 to 10, showing good agreement between the two equations.}
    \label{asymp}
\end{figure}

Moreover, we can show that this bound closely follows the asymptotic scaling of $ \mathcal{C}_{\rm max} (\bar{n}, N )$ with the number of beam splitters in the limit of large $N$, say, $N \gg \bar{n}$. This is because, in the limit of large $N$, we have, retaining just the lowest orders in $1/(N+1)$ 
\begin{equation}
S\left (\frac{\bar{n}}{N+1} \right ) = 1+\frac{\sqrt{\bar{n}}}{\sqrt{N+1}}+ \frac{\bar{n}}{\sqrt{2} (N+1)} + \ldots .
\end{equation}
We can use the convenient transformation
\begin{equation}
    \left [S \left (\frac{\bar{n}}{N+1} \right ) \right ]^{2(N+1)} = e^{2(N+1) \ln S (\bar{n}, N )} .
\end{equation}
Taking into account that
\begin{equation}
\ln (1+z) = z-\frac{z^2}{2}+  \ldots ,
\end{equation}
we have \begin{equation}
    2 (N+1) \ln S \left (\frac{\bar{n}}{N+1} \right ) = 2 \sqrt{\bar{n} (N+1)} + (\sqrt{2}-1) \bar{n} +  \ldots,
\end{equation}
so that coherence grows with $N$ asymptotically as 
\begin{equation}
\label{Casymp}
     \mathcal{C}_{\rm max} (\bar{n}, N ) \rightarrow e^{(\sqrt{2}-2) \bar{n}} e^{2 \sqrt{\bar{n}(N+1)}} -1 ,
\end{equation} 
which agrees well with the lower bound (\ref{lwG}). In Fig. \ref{asymp} we compare the asymptotic behavior in Eq. (\ref{Casymp}) with a numerical evaluation of Eq. (\ref{TC}) for $\bar{n}=1$ and $N$ ranging from 1 to 10, showing good agreement between them.

{}

\bigskip

Up to this point, we have shown that the energy has to be equally distributed between the output modes in order to maximize coherence. Thus, we can apply this consideration to any combination of beam splitters in order to optimize the resulting coherence. Here, we examine whether this optimal distribution can be achieved with the two different arrays of beam splitters shown in Figs. \ref{cascada1} and \ref{cascada2}.

For cascade 1 in Fig. \ref{cascada1}, the optimum balanced photon splitting can be achieved simply by considering that all beam splitters are balanced, having 50 \% reflectance and  50 \% transmittance. Concerning cascade 2 in Fig. \ref{cascada2}, the beam splitters can no longer all be 50 \%. Instead, we express the transmission and reflection coefficients of each beam splitter, which are assumed to be real without loss of generality, as
\begin{equation}
    t_j = \cos \theta_j, \quad r_j = \sin \theta_j, 
\end{equation}
for $j=1, \ldots, N$, where the beam splitters are numbered from left to right in Fig. \ref{cascada2}. 

For the last beam splitter on the right we must have 50 \% reflectance and  50 \%  transmittance, $t_N^2 = r_N^2$. For the second to last beam splitter, that is, $j=N-1$, the transmittance and reflectance must be in the ratio of 2:1 since we must feed two modes by transmission and one mode by reflection: that is, $t_{N-1}^2 = 2 r_{N-1}^2$. The general form of this series can easily be constructed to give the optimum configuration
\begin{equation}
 t_j^2 = \left ( N+1-j \right ) r_j^2  , 
\end{equation}
where $ N+1-j$ is the number of final output modes to the right of the $j$th beam splitter in cascade 2. In this way, the balance between transmission and reflection takes into account the number of final output modes that each beam splitter leads to. Finally, we can determine the angles $\theta_j$ as 
\begin{equation}
    \theta_j = \arcsin \left (\frac{1}{\sqrt{N+2-j}} \right ) .
\end{equation}

\bigskip

In this scenario where both the input and output are multimode coherent states, seemingly nothing would be changed by the beam splitter in regard to quantum properties. This will be the case if one restricts the analysis to the Glauber-Sudarshan quantum-optical theory of coherence. The Glauber-Sudarshan theory translates to the quantum domain the classical-optics theory of coherence by considering the correlations between the operator counterparts of the electric fields expressed in normal order. However, in this work we address a different theory, that is the quantum-mechanical coherence introduced by the pioneering works in Refs. \cite{JA06} and \cite{BCP14}. This theory is not restricted to electric fields, and does not work with field operators. Instead, it addresses the superposition of states in Hilbert spaces for arbitrary physical systems. Therefore, within the quantum-mechanical-coherence theory considered in this work, which is, indeed, different from the quantum-optical one, the  amount of quantum-mechanical coherence of Glauber-coherent states depends on the total intensity and its distribution between field modes, as we have just seen. 

To better understand our results in this quantum-mechanical context, we may regard coherence as being physically derived from phase correlations present in relative-phase statistics \cite{AL08,AL09,AL10}. This intuition is supported by the fact that maximum quantum-mechanical coherence for all coherence monotones holds for phaselike states \cite{BCP14}. In this regard, beam splitters produce pairs of modes with definite phase relations that can be used, for example, in interferometric arrangements, where one mode essentially provides a phase reference for the other mode experiencing the phase shift. This can be easily confirmed by the evaluation of the quantum phase difference between modes \cite{LS93,LS96,LP96b}. Before the beam splitter, the phase difference between the coherent state and the vacuum is fully random, which can be ascribed to the fully random phase of the vacuum. However, after the beam splitter the phase difference between the two modes is no longer fully random (cf. Refs. \cite{LS93,TTBSGAS00}). Actually, equal splitting is the case with smaller relative-phase uncertainty, in full agreement with the results reported above. Note that quantum-phase statistics is beyond the Glauber-Sudarshan quantum-optical theory, which cannot address quantum-phase observables. 

\begin{figure}[h]
    \includegraphics[width=6cm]{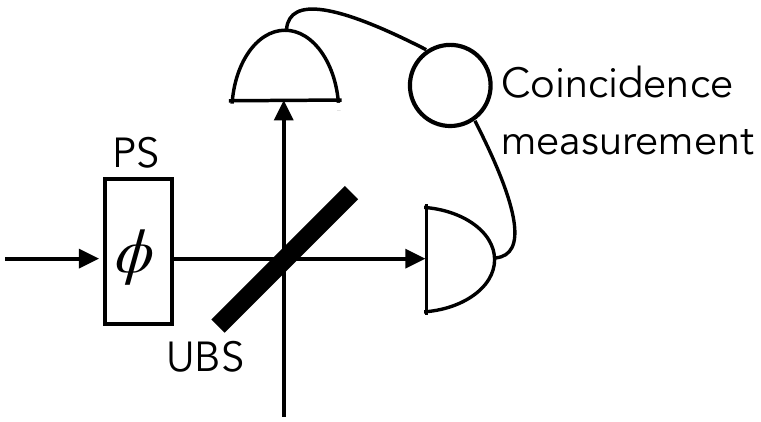}
    \caption{Scheme illustrating the quantum phase-difference measurement carried out in Ref. \cite{TTBSGAS00}.}
    \label{pm}
\end{figure}

In order to experimentally access our results we refer to the different strategies for measuring quantum-mechanical coherence. First, according to Eq. (\ref{fps}), the coherence of pure states is completely determined by the square root of the photon-number probabilities, which is a basic observable in quantum optics. For mixed states one can utilize quantum state tomography \cite{SBRF93,DLP95,WV96}, by which coherence can be determined once the density matrix of the state is retrieved. Tomography can be readily performed in multimode scenarios, which can actually be extended to the tomography of multimode input-output processes \cite{FFKL15}, including beam splitting \cite{LKKFSL08}. Regarding the measurement of the quantum phase difference invoked above, we refer to the experimental implementation already carried out in Ref. \cite{TTBSGAS00}. This measurement, illustrated in Fig. \ref{pm}, was carried out for the two-photon subspace, leveraging the fact that a suitable unbalanced beam splitter (UBS) transforms a phase-difference eigenstate, say, $|\phi =0 \rangle$, into the product of number states $|1\rangle |1\rangle $ at the output modes. After a phase-difference shift $\phi$ carried out by the phase shifter (PS), the coincidence detection of the two detectors at the two output ports of the unbalanced beam splitter gives the projection of the input state on any phase-difference eigenstate $|\phi \rangle$. Moreover, regarding quantum-phase relations, other formulations such as the one presented in Ref. \cite{LP93} can also be experimentally observed via double homodyne detection. Finally, practical ways of witnessing coherence can be found in Refs. \cite{WSRXCG21,WTWYKXLG17,YHTSXLG20}.

{}

\bigskip

\section{Number-state input}

In this  section we address the case of a highly nonclassical, pure, and incoherent input state, the number state $|n \rangle$, in one of the input modes, with the rest of the input modes again being in the vacuum state,
\begin{equation}
  |n ,0,\ldots,0 \rangle = \frac{1}{\sqrt{n!}}a^{\dagger n}_0 |0 ,0,\ldots,0 \rangle ,
\end{equation}
where $a_0$ is the complex-amplitude operator for the excited input mode. The output modes can be computed by the linear relation established by any cascade of beam splitters between the input and output modes, with $a_0$ expressed in terms of the output modes to determine that the output state is
\begin{equation}
\label{ss}
  \frac{1}{\sqrt{n!}} \left ( \sum_{j=0}^N \tau_j a_j^{\dagger} \right )^n |0,0,\ldots,0 \rangle,
\end{equation}
where $\tau_j$ are complex coefficients that must satisfy the following equality required to preserve commutation relations:
\begin{equation}
\label{coe}
    \sum_{j=0}^N |\tau_j|^2 = 1 .
\end{equation}
Without loss of generality, from now on we assume real coefficients $\tau_j$. After some simple algebraic manipulations via multinomial coefficients, we find that the output state expressed in the number basis is 
\begin{equation}
\label{ntau}
   |n, \tau \rangle =\sqrt{n!}  \sum_{\{m \}} \frac{\tau_0^{m_0}\tau_1^{m_1}\ldots \tau_N^{m_N}}{\sqrt{m_0 !m_1! \ldots m_N!}} |\{m \} \rangle ,
\end{equation}
where $\{m \}$ labels the collection of $N+1$ nonnegative integers respecting the photon-number conservation law
\begin{equation}
\label{pncl}
    m_0 + m_1 +\ldots +m_N = n .
\end{equation}
It is worth noting that these states are SU($N$+1) coherent states with a nice internal structure in terms of some kind of nesting of SU(2) coherent states \cite{KN00,LP96}. With this, the coherence becomes, in this case,
\begin{equation}
\label{Cnt}
    \mathcal{C} (n,N) = n! \left ( \sum_{\{m \}} \frac{\tau_0^{m_0}\tau_1^{m_1}\ldots \tau_N^{m_N}}{\sqrt{m_0 !m_1! \ldots m_N!}} \right )^2-1 .
\end{equation}
\bigskip

\bigskip

Now, we address the optimization of the coherence (\ref{Cnt}) when the coefficients $\tau_j$ are varied. We proceed by following the same {\it reductio ad absurdum} strategy assuming that the maximum of Eq. (\ref{Cnt}) is reached when at least two of the coefficients $\tau_j$ are not equal, say, $\tau_0 \neq \tau_1$, without loss of generality. We focus on the factor inside the parentheses to split the contribution of the photons within the  two assumed unequal modes,
\begin{equation}
   \sum_{\{m \}} \frac{\tau_0^{m_0}\tau_1^{m_1}\ldots \tau_N^{m_N}}{\sqrt{m_0 !m_1! \ldots m_N!}} = \sum_{k=0}^n \sum_{m=0}^k \frac{\tau_0^{m}\tau_1^{k-m}}{\sqrt{m! (k-m)!}} c_k ,
\end{equation}
where $c_k$ are the contributions of the rest of the modes that will share $n-k$ photons. For each $k$, the factor 
\begin{equation}
  \sum_{m=0}^k \frac{\tau_0^{m}\tau_1^{k-m}}{\sqrt{m! (k-m)!}} 
\end{equation}
corresponds to the coherence of the two-mode output state when a single beam splitter is illuminated by a number state $|k \rangle$ in one of the modes and a vacuum in the other mode. We also examined this case in a previous work \cite{AL23}, showing that for all $k$ the maximum value for such a factor holds for balanced splitting $\tau_0 = \tau_1$. 

This result can be deduced again intuitively from the fact that the factor is symmetric under the exchange of $\tau_0$ and $\tau_1$, so that $\tau_0 = \tau_1$ must be an extreme. That such an extreme is the maximum follows from the behavior of the binomial, so that the spreading of the photon-number distribution is clearly largest when $\tau_0 = \tau_1$, and the larger the spreading is, the larger the coherence is.

Therefore, the maximum coherence that can be obtained under these circumstances holds when all the $\tau_j$ coefficients are equal, that is,
\begin{equation}
\label{acae}
    \tau_j = \frac{1}{\sqrt{N+1}} ,
\end{equation}
which is the same equipartition of energy reached in the preceding section where the input was a Glauber coherent state. The explicit parameters and configurations for achieving (\ref{acae}) and thus the optimal coherence in this number-state case are then exactly the same as those found for the case of Glauber coherent states.

\bigskip

Therefore, 
\begin{equation}
\label{nmax}
   |n, \tau \rangle_{\rm max} =\sqrt{\frac{n!}{(N+1)^n}}  \sum_{\{m \}} \frac{1}{\sqrt{m_0 !m_1! \ldots m_N!}} |\{m \} \rangle ,
\end{equation}
and
\begin{equation}
\label{CSn}
    \mathcal{C}_{\rm max} (n,N)= \frac{n!}{(N+1)^n} \left ( \sum_{\{m \}} \frac{1}{\sqrt{m_0 !m_1! \ldots m_N!}} \right )^2-1 .
\end{equation}

\bigskip

\begin{figure}[h]
    \includegraphics[width=8cm]{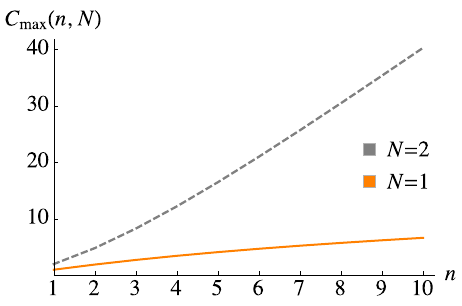}
     \caption{Plot of the maximum coherence for input number states $\mathcal{C}_{\rm max} (n,N)$ as a function of $n$ for a single beam splitter, $N=1$ (orange solid line), and a cascade of two beam splitters, $N=2$ (gray dashed line), showing the fast increase with the number of beam splitters and the number of photons.  }
    \label{numero}
\end{figure}

In Fig. \ref{numero}, we show the behavior of $\mathcal{C}_{\rm max} (n,N)$ as a function of $n$ for  a single beam splitter $N=1$ (orange solid line), and a cascade of two beam splitters, $N=2$ (gray dashed line), showing the increase with the number of beam splitters and the number of photons.  

\bigskip

Let us demonstrate analytically the limitless increase in coherence with the number of beam splitters $N$ following the same process as applied above to the case of coherent states. That is, using the multinomial theorem, 
\begin{equation}
  \sum_{\{m \}} \frac{1}{\sqrt{m_0 !m_1! \ldots m_N!}} \geq \sum_{\{m \}} \frac{1}{m_0 !m_1! \ldots m_N!} = \frac{(N+1)^{n}}{n!} ,
\end{equation}
and then 
\begin{equation}
\label{lwn}
    \mathcal{C}_{\rm max} (n,N) \geq  \frac{(N+1)^n}{n!}-1 ,
\end{equation}
proving the limitless increase in coherence with the number of beam splitters $N$.

{}

\bigskip

It might be interesting to compare $\mathcal{C}_{\rm max} (n,N)$ with the maximum coherence $C_{\mathrm{sup}}$ (it might be better to say supremum coherence) that might be reached in the finite-dimensional subspace that embraces the $n$ photons in $N + 1$ modes. This maximum is obtained for the phaselike state 
\begin{equation}
\label{maxs}
\frac{1}{\sqrt{D(n,N)}} \sum_{\{m \}} e^{i \phi (\{m \})} | \{m \} \rangle ,
\end{equation}
where $\phi (\{m \})$ are arbitrary phases and $D(n,N)$ is the dimension of the space spanned by the photon-number states $|\{m \} \rangle$, with $ m_0 + m_1 +\ldots +m_N = n$. This dimension can be easily computed as the number of different ways to distribute $n$ identical candies between $N+1$ distinguishable children, leading to the binomial coefficient
\begin{equation}
    D (n,N)  =  \begin{pmatrix} n+N\\N \end{pmatrix} .
\end{equation}
The supremum coherence is then,
\begin{equation}
\label{CSnmax}
    \mathcal{C}_{\mathrm{sup}} (n,N)= D(n,N) -1.
\end{equation}

\bigskip

\begin{figure}[h]
    \includegraphics[width=7.5cm]{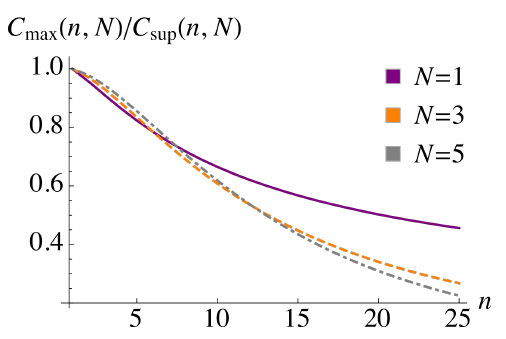}
    \caption{Plot of the quotient $\mathcal{C}_{\mathrm{max}} (n,N)/\mathcal{C}_{\mathrm{sup}} (n,N)$ as a function of $n$ for a single beam splitter, $N=1$ (purple solid line); three beam splitters, $N=3$ (orange dashed line); and five beam splitters, $N=5$ (gray dash-dotted line). }
    \label{maxsup}
\end{figure}

In Fig. \ref{maxsup} we plot the quotient $\mathcal{C}_{\mathrm{max}} (n,N)/\mathcal{C}_{\mathrm{sup}} (n,N)$ as a function of $n$ for a single beam splitter ($N=1$, purple solid line), three beam splitters ($N=3$, orange dashed line), and five beam splitters ($N=5$, gray dash-dotted line), showing a global departure from the supremum for increasing $n$. We can see that the single-photon case $n=1$ reaches the supremum for all $N$,
\begin{equation}
   C_{\mathrm{max}} (n=1,N)=C_{\mathrm{sup}} (n=1,N) .
\end{equation}

{}

\bigskip

\begin{figure}[h]
    \includegraphics[width=7.5cm]{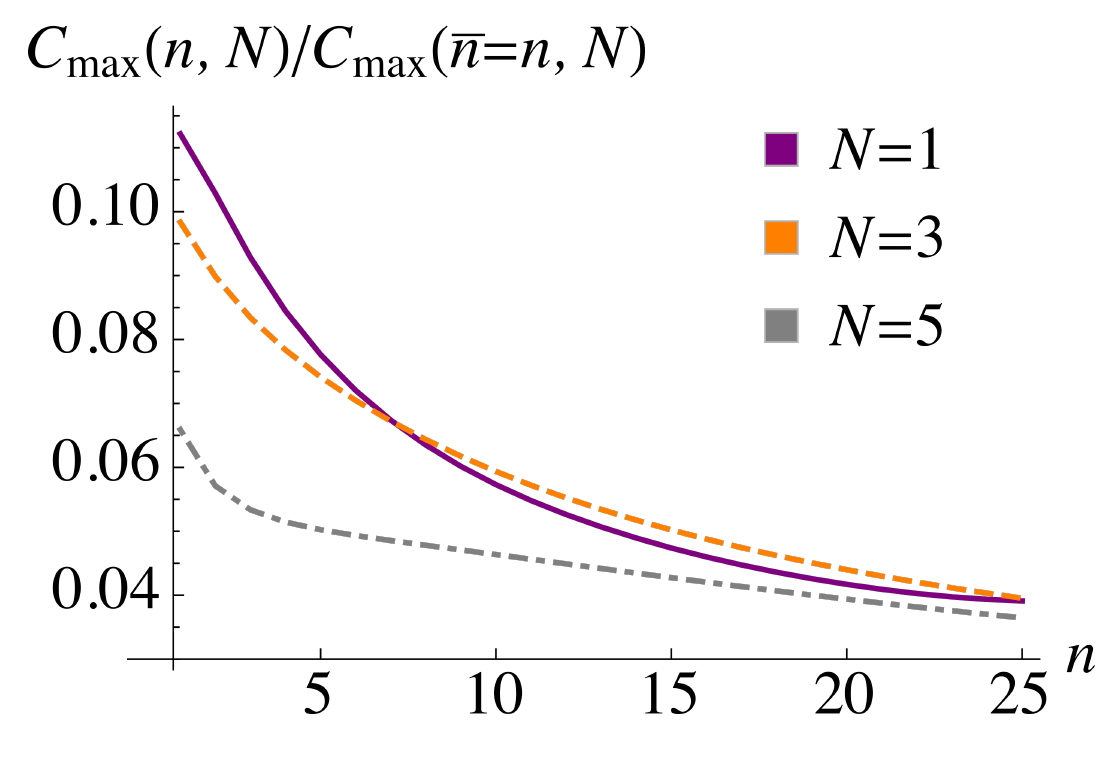}
    \caption{Plot of the quotient $\mathcal{C}_{\mathrm{max}} (n,N)/\mathcal{C}_{\mathrm{max}} (\overline{n}=n,N)$ of the optimum coherence for number states versus the optimum coherence for coherent states with the same mean number of photons, for one beam splitter ($N=1$, purple solid line), three beam splitters ($N=3$, orange dashed line) and five beam splitters ($N=5$, gray dash-dotted line). }
    \label{num/cohe}
\end{figure}

In Fig. \ref{num/cohe} we plot the quotient $\mathcal{C}_{\mathrm{max}} (n,N)/\mathcal{C}_{\mathrm{max}} (\overline{n}=n,N)$ of the optimum coherence for number states versus the optimum coherence for coherent states with the same mean number of photons for one beam splitter ($N=1$, purple solid line), three beam splitters ($N=3$, orange dashed line)  and five beam splitters ($N=5$, gray dash-dotted line). This shows that Glauber coherent states do much better than number states both for an increasing number of photons and for an increasing number of beam splitters, at least for the numerical ranges of parameters examined so far and in accordance with the corresponding bounds in Eqs. (\ref{lwG}) and (\ref{lwn}).

{}

\section{Mixed states}

After addressing the case for pure states, we examine the case of mixed states in the only excited input mode. More specifically, we consider two classes of mixed states. The first one is a statistical mixture of incoherent states, that is, diagonal in the number basis. The second one is classical-like, in the sense of the Glauber-Sudarshan $P$ function being a legitimate probability distribution. Both classes can be illustrated with the case of phase-averaged Glauber coherent states and thermal states, which belong to both categories. But we stress that there are incoherent states that are nonclassical, say, number states, and classical states that are partially coherent, say, Glauber coherent states different from the vacuum. 

Incoherent mixed states are diagonal in the number basis
\begin{equation}
\label{incoh}
    \rho = \sum_{n=0}^\infty p_n | n,0,\ldots,0 \rangle \langle n ,0,\ldots,0 |.
\end{equation}
For phase-averaged Glauber coherent states and thermal states the photon-number distributions are, respectively, 
\begin{equation}
    p_{\mathrm{pa},n} = \frac{\bar{n}^n}{n!}e^{-\bar{n}}, \quad p_{\mathrm{th},n} = \frac{1}{\bar{n}+1}\left ( \frac{\bar{n}}{\bar{n}+1} \right )^n .
\end{equation}

In the general case, the output state will take the form 
\begin{equation}
  \rho_\tau = \sum_{n=0}^\infty p_n | n, \tau \rangle \langle n, \tau |,
\end{equation}
where $| n, \tau \rangle$ are the states in Eq. (\ref{ntau}). Therefore, the coherence can be computed as
\begin{equation}
    \mathcal{C} \left ( p ,N \right )= \sum_{n=0}^\infty p_n \sum_{\{ m \},\{ m^\prime \}} |c^{(n)}_{\{ m \}}||c^{(n)}_{\{ m^\prime \}}|-1 ,
\end{equation}
which is equivalent to
\begin{equation}
    \mathcal{C} \left ( p ,N \right )=\sum_{n=0}^\infty p_n \left ( \sum_{\{ m \}} |c^{(n)}_{\{ m \}}|\right )^2 -1 ,
\end{equation}
where $c^{(n)}_{\{ m \}}$ are the photon-number coefficients in Eq. (\ref{ntau}).

Finally, resorting to the coherence of the pure photon-number case $\mathcal{C} (n,N)$, we get
\begin{equation}
\label{aCn}
    \mathcal{C} \left ( p ,N \right ) = \sum_{n=0}^\infty p_n \mathcal{C} (n,N) = \overline{\mathcal{C} (n,N)} ,
\end{equation}
where the average denoted by the overline is taken with respect to the photon-number distribution $p_n$. So the coherence of the output state is the average of the coherences obtained with each photon-number state, which respects the convexity condition of coherence measures \cite{BCP14}. 

Finally, it is clear that the maximum coherence will be obtained when the beam splitters are arranged to give the equipartition in Eq. (\ref{acae}) since it does not depend on the number of photons $n$, that is,
\begin{equation}
    \mathcal{C}_{\rm max} \left ( p ,N \right ) =\overline{\mathcal{C}_{\rm max} (n,N)} .
\end{equation}
We stress that this holds for every state of the form (\ref{incoh}). 

\bigskip

There is another approach to the mixed case in terms of the $P$ function of Glauber and Sudarshan \cite{MW95},
\begin{equation}
\label{class}
    \rho = \int d^2 \alpha P(\alpha) | \alpha,0,\ldots,0 \rangle \langle \alpha ,0,\ldots,0 | ,
\end{equation}
where we will always consider classical-like states with $P(\alpha)$ functions that are legitimate probability distributions on the complex plane. For phase-averaged Glauber coherent and thermal states we have, respectively,
\begin{equation}
    P_{\rm pa}(\alpha ) = \delta \left ( |\alpha |^2 = \bar{n} \right ), \qquad  P_{\rm th} (\alpha ) = \frac{1}{\pi \bar{n} } e^{-|\alpha |^2/\bar{n}} .
\end{equation}
In the optimum case the output state $\rho_\tau$ will take the form  
\begin{equation}
    \rho_\tau = \int d^2 \alpha  P(\alpha) \bigotimes_{j=0}^N \left | \frac{\alpha}{\sqrt{N+1}}  \right\rangle_j  \left\langle \frac{\alpha}{\sqrt{N+1}} \right | .
\end{equation}
Following essentially the same steps as above, 
\begin{equation}
    \mathcal{C}_{\rm max} \left ( P ,N\right ) = \int d^2 \alpha P(\alpha)  \mathcal{C}_{\rm max}\left ( |\alpha |^2, N \right )  ,
\end{equation}
where $\mathcal{C}_{\rm max}\left ( |\alpha |^2, N \right )$ is the coherence in Eq. (\ref{TC}) for a pure coherent state with a mean number of photons $\bar{n} = |\alpha |^2$. The final result  can again be expressed as an average,
\begin{equation}
\label{aCa}
    \mathcal{C}_{\rm max} \left ( P ,N \right ) =\overline{\mathcal{C}_{\rm max} (|\alpha|^2,N)},
\end{equation}
where the average is with respect to $P(\alpha)$. We stress that this holds for every classical-like state of the form (\ref{class}) with {\it bona fide} $P(\alpha)$ as s probability distribution. 

\bigskip

As a straightforward conclusion of Eqs. (\ref{aCn}) and (\ref{aCa}) we find that the output coherence when the input is a phase-averaged Glauber coherent state, which is an incoherent state, is exactly the same as that for a pure Glauber coherent input state of the same mean photon number. 

\bigskip

Regarding thermal states, 
\begin{equation}
    \mathcal{C}_{\rm max} \left ( P_{\rm th} ,N \right ) =\frac{1}{\pi \bar{n}} \int d^2 \alpha e^{-|\alpha |^2/\bar{n}} \mathcal{C}_{\rm max} (|\alpha|^2,N) ,
\end{equation}
we may proceed further by considering a large mean number of photons so that the approximation (\ref{CTapp}) can be safely used, leading to 
\begin{equation}
    \mathcal{C}_{\rm max} (|\alpha|^2,N) \simeq  \left ( \frac{8\pi}{N+1} \right ) ^\frac{N+1}{2} |\alpha|^{N+1} -1 ,
\end{equation}
and then, 
\begin{equation}
\label{Ctapp}
    \mathcal{C}_{\rm max} \left ( P_{\rm th} ,N \right ) =\Gamma \left ( \frac{N+3}{2} \right ) \left ( 8\pi \frac{\bar{n}}{N+1} \right )^\frac{N+1}{2}  -1 ,
\end{equation}
where $\Gamma$ is the gamma function. 

\bigskip

Comparing Eqs. (\ref{CTapp}) and (\ref{Ctapp}), we can note that they differ only in the presence of the $\Gamma$ prefactor in the thermal case, which makes coherence grow much faster with $N$ than it does for Glauber coherent states with the same mean photon number. This is an effect of nonlinearity in the dependence of coherence on intensity, which, as we have already noted, is a rather novel feature in coherence theories. 

\section{Conclusions}

We saw that coherence can grow without limit by increasing the number of beam splitters in all the scenarios considered here. These devices are usually considered rather passive in the classical optical scenario, so it is interesting that they may easily increase such a valuable quantum resource in its quantum counterpart. Actually, layouts akin to those examined here are the first step in all interferometric protocols precisely in this spirit.

We showed that, in this coherence production, the initial coherence of the input state plays no role, as demonstrated by the phase-averaged states, that are incoherent but give rise to the same coherence as the initial coherent states. Furthermore, classical incoherent input thermal states produce larger quantum coherence than Glauber coherent states with the same mean photon number, which seems to defy standard intuition. 

Moreover, rather paradoxically, quantumness seems to play no role since strongly quantum states, such as the photon-number states, lead to a much smaller amount of quantum coherence than classical-like states with the same mean number of photons. 

In this work, utilizing states fully characterized by the mean number of photons allows us to compare the increase in coherence for very different states with regard to photon statistics. In this regard, the analysis can be extended to squeezed states, utilizing the squeezing parameter as an additional degree of freedom in the optimization process.
For states with more exotic photon statistics, such as NOON states, a variable other than the mean number of photons would be needed in order to compare the results.

Finally, note that the analysis of configuration 1 could be extended to investigate the impact of varying coherence due to the bias of the coin in a quantum walk scenario.

\bigskip

\noindent{\bf Acknowledgments.- }
L. A. and A. L. acknowledge financial support from Project No.PR44/21--29926 of Santander Bank and Universidad Complutense de Madrid. The authors would like to thank the anonymous referee for valuable suggestions that helped  to improve the manuscript.

\end{document}